\documentstyle[11pt,paspconf,epsf]{article}
\begin{document}
\title{Gravitational lensing studies with the 4-m International Liquid
Mirror Telescope (ILMT)}
\author{C. Jean, J.-F. Claeskens\altaffilmark{1} and J.
Surdej\altaffilmark{2}}
\affil{Institut d'Astrophysique et de G\'eophysique, Universit\'e de
Li\`ege \,
Avenue de Cointe 5, B--4000 Li\`ege, Belgium}
\altaffiltext{1}{Also Charg\'e de Recherches du Fonds National de la
Recherche Scientifique, Belgium}
\altaffiltext{2}{Also Directeur de Recherches du Fonds National de la
Recherche Scientifique, Belgium}

\begin{abstract}
A 4\,m International Liquid Mirror Telescope (ILMT) is being built in the 
north of Chile by an international consortium and will become operational
in two years from now. We present here a short
description of the telescope as well as estimates of the
microlensing, macrolensing and weak lensing effects expected from
a deep, multicolor imaging survey made with such a telescope.
\end{abstract}

\section{What is a Liquid Mirror Telescope?}

A Liquid Mirror Telescope (hereafter LMT) consists of a container, filled
with mercury, which spins around a vertical axis at a constant speed. Thus,
the surface of the reflecting liquid takes the shape of a paraboloid which
can be used as the primary mirror of a telescope. By placing a CCD detector
at the prime focus of the mirror, one obtains a telescope suitable for
astronomical observations. Because an LMT cannot be tilted and hence cannot
track
like conventional telescopes do, the time delay integration (TDI) technique
(also known as drift scan) is used to collect the light of the objects
during their transit along the CCD detector. A semi-classical corrector is
added in front
of the CCD detector in order to provide a larger field of view and to remove
the TDI distortion. This distortion arises because the
images in the focal plane move at different speeds on distinct curved
trajectories while the TDI technique moves
the pixels on the CCD at a constant speed along a straight line.

\section{The 4\,m ILMT}

The 4\,m International Liquid Mirror Telescope (ILMT) will be installed in
the Atacama desert in Chile and will be fully dedicated
to a zenithal direct imaging survey in two
broad spectral bands (B and R). The possible construction
of an array of several ($\geq$ 2) liquid mirrors, working at different
wavelengths, is also being considered. It should allow one to reach
limiting magnitudes B = 23.5 and R = 23 in a single scan
of a $4096\times4096$ pixels CCD or an equivalent mosaic of four
$2048\times2048$ pixels CCDs.
The telescope field of view is about 30 x 30 arcminutes and the telescope 
will be operated
during no less than 4-5 years.  Thus, very precise photometric and
astrometric data will be obtained in the drift scan mode night after
night, during several consecutive months each year, for all objects
contained in a strip of sky of approximately 140 square degrees, at
constant declination. Due to its location in the Atacama desert, both
low and high galactic latitude regions will be studied. The low
galactic latitudes are propitious to microlensing effects and the high
galactic latitudes to observations of macrolensing by galaxies as
well as to strong and weak lensing effects induced by galaxy
clusters.

\section{Gravitational lensing studies}

Numerical simulations were carried out to estimate the gravitational
lensing effects we can expect from a survey made with a 4\,m LMT.
\subsection{Microlensing in the Galaxy}
We used a galactic model with 3 components: the halo, the disk and the
Galactic bulge.
About 50 (resp. 10, 3) microlensing events due to the bulge (resp. the
disk, the halo) are expected after one year of ILMT observations at an
observing latitude of $-29^\circ$ assuming that the
Galaxy is entirely made of 1 M$_\odot$ dark compact objects.
\subsection{Macrolensing}
Considering the quasar number counts relation and the optical depth of
cosmologically distributed ``singular isothermal spherical'' galaxies,
we expect to detect approximately 50 new multiply imaged quasars.
\subsection{Weak lensing}
We used a model similar to that described by Nemiroff and Dekel (1989, ApJ,
344, 51) and conclude that
in a survey of $100^{\circ^2}$ with a limiting brightness
$\textrm{B}_{\textrm{lim}}$ of 26.5 mag/arcsec$^2$ or fainter, one can expect
at least 50 luminous arcs (axial ratio $A\geq5$, angular extent $\theta\geq10^{\circ}$).
\subsection{Monitoring and determination of H$_0$}
The daily monitoring of the 50 new lensed quasars will significantly
contribute to a statistical and independent determination of the Hubble
constant and to a better understanding of the QSO source structure and of the
distribution of dark matter in the Universe, through the analysis of
microlensing effects.

\vspace{1cm}
\noindent
\textit{Acknowledgments: }
It is a pleasure to thank Martin Cohen for his kind help and his calculations of star counts with his SKY program. We also thank Annie Robin and her team for making their galactic model available through the WEB.

\end{document}